\documentclass[prb,twocolumn,aps,superscriptaddress,a4paper,showpacs]{revtex4}

\usepackage{dcolumn}
\usepackage{amsmath}
\usepackage{graphicx}
\usepackage{latexsym}
\usepackage{amsfonts}
\usepackage{amssymb}

\newcommand{\be}{\begin{equation}}
\newcommand{\ee}{\end{equation}}
\newcommand{\bea}{\begin{eqnarray}}
\newcommand{\eea}{\end{eqnarray}}

\newcommand{\Tr}{ {\rm Tr} \, }

\begin{document}


\def\a{\alpha}
\def\b{\beta}
\def\g{\gamma}
\def\G{\Gamma}
\def\d{\delta}
\def\D{\Delta}
\def\e{\epsilon}
\def\ve{\varepsilon}
\def\z{\zeta}
\def\h{\eta}
\def\th{\theta}
\def\k{\kappa}
\def\l{\lambda}
\def\L{\Lambda}
\def\m{\mu}
\def\n{\nu}
\def\x{\xi}
\def\X{\Xi}
\def\p{\pi}
\def\P{\Pi}
\def\r{\rho}
\def\s{\sigma}
\def\S{\Sigma}
\def\t{\tau}
\def\f{\phi}
\def\vf{\varphi}
\def\F{\Phi}
\def\c{\chi}
\def\w{\omega}
\def\W{\Omega}
\def\Q{\Psi}
\def\q{\psi}


\def\de{\partial}
\def\inf{\infty}
\def\ra{\rightarrow}
\def\bra{\langle}
\def\ket{\rangle}
\def\grad{\mbox{\boldmath $\nabla$}}
\def\ua{\uparrow}
\def\da{\downarrow}
\def\di{{\rm d}}
\def\xc{{\rm xc}}
\def\Tr{{\rm Tr}\,}
\def\tc{\tilde{\chi}}
\def\tA{\tilde{A}}
\def\bu{\bar{1}}
\def\bd{\bar{2}}
\def\bt{\bar{3}}
\def\bq{\bar{4}}


\def\bj{\mbox{\boldmath $j$}}
\def\bbf{\mbox{\boldmath $f$}}
\def\bg{\mbox{\boldmath $g$}}
\def\bk{\mbox{\boldmath $k$}}
\def\bp{\mbox{\boldmath $p$}}
\def\bbq{\mbox{\boldmath $q$}}
\def\br{\mbox{\boldmath $r$}}
\def\bv{\mbox{\boldmath $v$}}
\def\bA{\mbox{\boldmath $A$}}
\def\bB{\mbox{\boldmath $B$}}
\def\bE{\mbox{\boldmath $E$}}
\def\bF{\mbox{\boldmath $F$}}
\def\bG{\mbox{\boldmath $G$}}
\def\bJ{\mbox{\boldmath $J$}}
\def\bL{\mbox{\boldmath $L$}}
\def\bM{\mbox{\boldmath $M$}}
\def\bP{\mbox{\boldmath $P$}}
\def\bL{\mbox{\boldmath $\Lambda$}}
\def\bS{\mbox{\boldmath $\Sigma$}}
\def\bal{\mbox{\boldmath $\alpha$}}
\def\bdel{\mbox{\boldmath $\delta$}}

\title{Conserving Approximations in Time-Dependent Density Functional
Theory}

\author {Ulf von Barth}
\affiliation{Solid State Theory, Institute of Physics, Lund University,
S\"olvegatan 14 A, S-22362 Lund, Sweden}
\author {Nils Erik Dahlen}
\affiliation{Rijkuniversiteit Groningen, Theoretical Chemistry,
Materials Science Center, 9747AG, Nijenborgh 4, Groningen, The
Netherlands}
\author {Robert van Leeuwen}
\affiliation{Rijkuniversiteit Groningen, Theoretical Chemistry,
Materials Science Center, 9747AG, Nijenborgh 4, Groningen, The
Netherlands}
\author {Gianluca Stefanucci}
\affiliation{Solid State Theory, Institute of Physics, Lund University, 
S\"olvegatan 14 A, S-22362 Lund, Sweden}

\date{\today}

\begin{abstract}
In the present work we propose a theory for obtaining successively
better approximations to the linear response functions of time-dependent
density or current-density functional theory. The new technique is based
on the variational approach to many-body perturbation theory (MBPT) as
developed during the sixties and later expanded by us in the mid
nineties. Due to this feature the resulting response functions obey
a large number of conservation laws such as particle and momentum
conservation and sum rules. The quality of the obtained results is
governed by the physical processes built in through MBPT but also
by the choice of variational expressions. We here present several
conserving response functions of different sophistication to be used
in the calculation of the optical response of solids and nano-scale
systems.
\end{abstract}


\maketitle

\section{Introduction}

Optical spectra constitute important tools for gaining information
on the electronic structure of solids, molecules, and nano-systems.
In many systems the particle-hole interaction leads to a strong excitonic
distortion of the optical spectrum - particularly in nano-scale objects.
The theoretical description of such spectra is relatively sophisticated
and very costly from a computational point of view. Some time ago,
it was realized that these spectra are also within reach using 
time-dependent (TD)
density-functional theory (DFT) - but with much less computational 
effort. From the Runge-Gross theorem\cite{rg0} of TDDFT we know how to
construct the exact density response function of any electronic system
in terms of an exchange-correlation kernel describing the particle-hole
interactions. And from recent work by several researchers\cite{ro,mr} 
we have a rather good idea about the properties of
this kernel if it is to reproduce the rather accurate results obtained
from solving the Bethe-Salpeter equation of many-body perturbation theory
(MBPT). The kernel, usually named $f_{\xc}$, has been calculated in the
exchange-only approximation of TDDFT by several people in the past,
see for instance Refs. \onlinecite{gt1,has}. More recently, the 
kernel $f_{\xc}$ has been calculated in the same approximation by 
Petersilka, Gossmann, and Gross \cite{pgg} 
for the helium atom, by Kurth and von Barth for the density
response of the homogeneous electron gas,\cite{kvB} and by
Kim and G\"orling\cite{kg} in the case of bulk silicon.
In the cases of atomic helium and the homogeneous electron gas the 
resulting response function represented a substantial improvement on
that of the Random Phase Approximation (RPA). 
The excitation energies of helium were much improved
and the total energies obtained from the response function were much 
superior to those obtained from the RPA response function in both
helium and the homogeneous electron gas.  Unfortunately, this
{\em ab initio} approach did not work very well in bulk silicon
unless one rather arbitrarily introduces some kind of static screening
of the particle-hole interaction.

In actual fact, within TDDFT, no systematic and realistic route toward
successively better approximations has, so far, been available. In the
present work we have constructed such a scheme based on the 
variational approach to many-body theory developed in Ref.\onlinecite{abl}.
In terms of the one-electron Green function of MBPT, these
functionals give stationary expressions for the total action of the
system at hand - or the total energy in the case of time independent
problems. From a stationary action it is rather straight-forward to
construct the time dependent density response function. Building the
functionals from the $\F$-derivable theory of Baym and
Kadanoff,\cite{baym,bk} always
results in response functions which obey essential physical
constraints like particle, momentum or energy conservation.

The simple idea of the present work is to restrict the variational
freedom of the functionals to the domain of Green functions which are
non-interacting and given by a local one-electron potential - and 
vector-potential in case of current-DFT. According
to the Runge-Gross theorem this restriction immediately results in a
density-functional theory the quality of which is determined
by the sophistication which is build in to the choice of $\F$ derivable
approximation for the action functional. Thus, to every conserving scheme
within MBPT there is a corresponding level of approximation within TDDFT.
The latter is determined variationally and there is no longer a need
for an {\em ad hoc} procedure to equate corresponding quantities between
TDDFT and MBPT. A potentially interesting consequence of the theory
proposed here, is that the often discussed linearized Sham-Schl\"uter 
equation\cite{ss} for the exchange-correlation potential is nothing but the 
stationary condition for the action functional.
In the particular version of the variational functionals
developed in Ref. \onlinecite{abl} and named $\Q$ derivable theories, also the screened
Coulomb interaction becomes an independent variable at ones disposal.
This leads to approximations within TDDFT which are potentially as
accurate as those of more elaborate schemes within MBPT but which are
comparatively easier to implement - especially in nano-systems and
complex solids.

\section{Variational Approach to TDDFT}

Let us consider a system of interacting fermions exposed to an 
external, possibly time-dependent field $w(\br t)$. The full 
many-body Hamiltonian reads  
\begin{equation}
\hat{H}=\hat{T}+\hat{U}+\hat{W},
\label{fham}
\end{equation}
where 
$$
\hat{T}=-\frac{1}{2}\int d^3 r \; \q^{\dag}(\br)\nabla^{2}\q(\br)
$$
is the kinetic operator, while
$$
\hat{U}=\frac{1}{2}\int d^3 r d^3 r' \;\q^{\dag}(\br)\q^{\dag}(\br')v(\br,\br')
\q(\br')\q(\br),
$$
is the interaction operator ($v(\br,\br')=1/|\br - \br'|$).
The coupling to the external field is given by
$$
\hat{W}=\int d^3 r \;w(\br t) \, \hat{n}(\br),
$$
where $\hat{n}(\br)=\q^{\dag}(\br)\q(\br)$ is the density
operator. The Green function $G$ obeys Dyson's equation
$$
G=G_{\rm H}+G_{\rm H}\S G
$$
where $G_{\rm H}$ is the Hartree Green function and $\S$ is the
exchange-correlation part of the electronic self-energy. Diagrammatic
perturbation theory provides a tool for
generating approximate self-energies and, in turn, approximate Green
functions. Except for physical intuition, the diagrammatic techniques
rely solely on the validity of Wick's theorem.\cite{fw, d84} Thus,
a typical contribution to the self-energy is represented by a
diagram containing non-interacting propagators and interaction lines.
However, any approximation which contains only a finite number
of these diagrams violates many conservation laws. Conserving
approximations require a proper choice of an infinite set of diagrams.
The conserving approach by Baym\cite{baym} was based on such choices.
Also the variational scheme by Almbladh, von Barth and van
Leeuwen\cite{abl} (ABL) was designed with the same objective in mind.
The former approach is referred to as a $\F$-derivable scheme
because its central quantity is a universal functional, called $\F$, 
of the one-electron Green function $G$ and the bare Coulomb potential
$v$. It is constructed such that its functional derivative with
respect to $G$ gives the exchange-correlation part of the electronic
self-energy $\S$ whereas the functional derivative with respect to the
Coulomb interaction $v$ essentially gives the reducible polarizability
$\chi$ of the system,
\begin{equation}
\S(1,2)=\frac{\d \F}{\d G(2,1)}; \quad
\chi(1,2)= -2 \frac{\d \F}{\d v(2,1)} \, .
\label{baymse}
\end{equation}
(Here and in the following we use the short-hand notation 
$1=(\br_{1},t_{1})$, $2=(\br_{2},t_{2})$ and so on).
Notice, however, that there is no reference to an actual
system in the $\F$ functional. It acquires a meaning only when it is 
evaluated at a Green function of an actual system.
In the approach of ABL, the central quantity is instead the 
functional $\Q$ having the Green function $G$ and the screened
Coulomb interaction $W$ as independent variables. It is constructed
so as to give the self-energy when it is differentiated with respect to
$G$ and the irreducible polarizabillty $P$ when functionally
differentiated with respect to $W$. Again, there is no reference to
the actual system contained in the functional $\Q$.
By adding functional pieces to the $\F$ or the $\Q$ functional
respectively, pieces which do contain clear connections to the
system under study (like, {\em e.g.}, the externally applied potential
$w$), one constructs functionals for
the total energy - or the action in the case of time dependent problems -
which, as functionals of $G$, have their stationary point at the Green
function $G$ which is the solution to Dyson's equation. In the case
of the $\Q$-based functionals they are also stationary when the
screened interaction $W$ obeys the so called reduced Bethe-Salpeter
equation to be discussed later.

The first variational functional of this kind was constructed by
Luttinger and Ward\cite{lw} (LW). It is a $\F$ functional and it has
the appearance
\begin{equation}
i Y_{\rm LW}[G]=\F[G]-\Tr\left\{\S G+\ln(\S-G_{\rm H}^{-1})\right\}-
i U_{\rm H}[G].
\label{lwf}
\end{equation}
In Eq. (\ref{lwf}), the functional
$U_{\rm H}[G]= - \frac{i}{2}\Tr\{V_{\rm H}G\}$ is the classical Hartree
energy, $V_{\rm H}({\bf r} t)=\int d^{3}r' v({\bf r},{\bf r}')n({\bf r}' t)$ 
is the Hartree potential
and $n(\br t)$ the electron density.
The symbol $\Tr$ (Trace) denotes a sum over labels of one-electron
states plus an integration over time, or frequency for equilibrium
problems in the ground state, a sum over discrete frequencies or an
integral over imaginary times for elevated temperatures or an integral
along the Keldysh contour~\cite{Keldysh:65,d84,Varenna:04} in
the case of non-equilibrium problems.\cite{rg}
It is straightforward to realize that $Y_{\rm LW}$ is stationary
when $G$ obeys Dyson's equation with the self-energy of Eq.
(\ref{baymse}).
At the stationary point, the Green function is fully conserving.

At this point we would like to draw the readers attention to 
a very interesting fact, the
ramifications of which have yet to be discovered. The variational
schemes are, by no means, unique. By adding to
$Y_{\rm LW}$, any functional $F[D]$, where
\be
D[G]=G(G_{\rm H}^{-1}-\S[G])-1,
\ee
obeying
$$
F[D=0] = {\left(\frac{\d F}{\d D}\right)}_{D=0} = 0,
$$
one obtains a new variational functional having the same stationary 
point and the same value at the stationary point. It might, however, be 
designed to give 
a second derivative which also vanishes at the stationary point - 
something that would be of utmost practical value. Such possibilities
could open up a whole new field of research.

Choosing to add $F[D]$ to the LW functional, where
\be
F[D]=\Tr\left\{-D+\ln(D+1)\right\}
\label{fdfunc}
\ee
obviously has the desired properties, leads to the functional
$$
 i Y_{\rm K}[G]=\F[G]-\Tr\left\{
GG_{\rm H}^{-1}-1+\ln(-G^{-1})\right\} - i U_{\rm H}[G].
$$
This functional was first written down by Klein,\cite{klein}
and could thus be
called the Klein functional in order to distinguish it from the LW
functional above. Unfortunately, this functional is less stable 
(large second derivative) at the stationary point as compared to the 
LW functional.\cite{dvB,dvBMP2,dlvBIJQC,dlvB} Since the construction of response functions
for TDDFT from the variational functionals involve evaluating them
at non-interacting Kohn-Sham Green functions, one might expect a less
stable functional to give rise to inferior response functions. And this
is something which has to be thoroughly investigated. But it is clear
that the Klein functional is much easier to evaluate and manipulate
as compared to, {\em e.g.}, the LW functional.

All the $\F$ functionals lead to a Dyson equation which has to be 
solved self-consistently for $G$. This is, in general, a very demanding 
task because of the complicated satellite structure inherent to any
interacting Green function. This severe complication is, however,
circumvented by switching to TDDFT.

Our approximations within TDDFT are just special cases of the
variational functionals in which
we restrict the variational domain of the Green function to be all
Green functions obtainable from a one-electron Schr\"odinger equation
with a {\em local} multiplicative potential - or vector potential in
the case of current-DFT.

We remark that this restriction on the variational freedom renders
all the variational functionals density functionals.\cite{rg,hk,rg0}
Given a density
there is a local potential which in a non-interacting system produces
that density. This potential produces the non-interacting Green
function which we use to evaluate our functionals. Thus, the
variational approach naturally generates different approximations
within DFT for static problems and within TDDFT for time-dependent
problems or for the response functions of stationary problems.
As we shall see, the exchange-correlation quantities depend on the 
choice of the action functional so that to every approximate Baym functional $\F$ 
correspond different approximate exchange-correlation potentials and 
kernels.

Below, we discuss TDDFT and TD current-DFT (TDCDFT)
approximations in the framework of the Klein functional
and of the LW functional. We also generalize the theory to $\Q$
functionals and give some examples of approximations which we
believe to be quite feasible to apply to realistic systems
taking due account of the full electronic structure of one-body
origin.

\section{TDDFT from the Klein Functional}
\label{dftkf}

Let $G_{s}$ be the Green function of a non-interacting system of
electrons exposed to the external, possibly time-dependent, potential
$V(\br t)$. The Klein functional evaluated at $G_{s}$ can then be
regarded as a functional of $V$:
$$
i Y_{\rm K}[V]=\F[G_{s}]-\Tr\left\{
G_{s}G_{\rm H}^{-1}-1+\ln(-G_{s}^{-1})\right\}- i U_{\rm H}[G_{s}].
$$
We could now directly use the stationary property of the Klein
functional with respect to variations in the unknown one-body potential
$V$ in order to obtain an equation for that potential.
Because of the simplicity of a noninteracting Green function, however,
the functional
$Y_{\rm K}$ can first be manipulated to acquire a physically appealing
form. This can, most easily, be seen in the static case elaborated below.
The following equations are still valid in the case of time dependent
problems and/or problems at elevated temperatures. This, however,
requires some reinterpretations of standard DFT quantities like,
{\em e.g.}, $T_s$ or $U_{\rm H}$, which then become functionals on the
Keldysh contour.\cite{keldtut}
For non-interacting Green functions the logarithm of the inverse
of $G_s$ is just the sum of the occupied eigenvalues contained in $G_s$.
\cite{lw}
And the trace of $G_{s}G_{\rm H}^{-1}-1$ is just the integral of the
particle density multiplied by the potentials $V - w - V_{\rm H}$.
Expressing the eigenvalues of the one-electron Hamiltonian $-\nabla^2/2+V$
as expectation values then leads to,
\be
Y_{\rm K}[V]= -i \F[G_{s}] + T_s[n] + \int w n + U_{\rm H} .
\label{kleinxc}
\ee
Here, the quantity $T_s[n]$ is the well known functional for the
kinetic energy of non-interacting electrons in the potential $V$ - which
produces the density $n$. Comparing now with standard DFT we
see that the $\F$-functional precisely plays the role of the
exchange-correlation energy. This means that we may reuse standard
DFT results and realize that the Klein functional is stationary when
\be
 V = w + V_{\rm H} - i \frac{\d \F}{\d n}.
\ee
where the last term is the exchange-correlation potential $v_{\xc}$.
Using the chain rule for differentiation we can rewrite $v_{\xc}$ as
\be
v_{\xc} = - i \frac{\d \F}{\d n}
  = - i\int \frac{\d \F}{\d G_s} \frac{\d G_s }{\d V } \frac{\d V}{\d n}
\ee
Remembering that the derivative of $\F$ with respect to the Green function
is just the self-energy $\S$ of our $\F$ derivable theory and that the
last factor is the inverse of the density response function $\chi_s$ of
non-interacting electrons, we finally arrive at the equation

\begin{equation}
\int
\S_{s}(2,3)\L(3,2;1) \; d(23) =
\int \chi_{s}(1,2)v_{\xc}(2) \; d2,
\label{lss}
\end{equation}
where we have defined a generalized non-interacting response function $\L$
according to
$$
i\L(2,3;1)\equiv
\frac{\d G_{s}(2,3)}{\d V(1)}=G_{s}(2,1)G_{s}(1,3),
$$
in terms of which we have $\chi_{s}(1,2)=\L(1,1;2)$.

Eq. (\ref{lss}) is exactly the ``linearized'' form of the
Sham-Schl\"uter (SS) equation, {\em i.e.}, it can be obtained from
the SS equation \cite{ss} by replacing the interacting $G$ with $G_{s}$, and
$\S$ with $\S_{s}=\S [G_{s}]$.
Thus, the ``linearized'' SS equation
follows from a variational principle. We realize that the variational
approach can be used to obtain
successively better approximations to the exchange-correlation
potential $v_{\xc}$ by making successively better approximations to the
functional $\F$. In addition, the $\F$- derivability and the variational
property renders any approximation fully conserving.

The full density response function expressed in the manner of
TDDFT is\cite{pgg}
$$
\chi=\chi_{s}+\chi_{s}(v+f_{\xc})\chi, \quad
f_{\xc}(1,2)=\frac{\d v_{\xc}(1)}{\d n(2)}.
$$
The kernel $f_\xc$ can now be obtained from one further variation with
respect to the total
potential $V$. The variation of $v_{\xc}$ with respect to $V$ can be
expressed in terms of the exchange-correlation kernel $f_\xc$ as
$$
\frac{\d v_{\xc}(1)}{\d V(2)}=\int f_{\xc}(1,3)\chi_{s}(3,2)\; d3.
$$
Thus, by varying Eq. (\ref{lss}) we obtain the following equation for $f_{\xc}$
\begin{eqnarray}
\int \chi_{s}(1,3)f_{\xc}(3,4)\chi_{s}(4,2)\; d(34) \nonumber \\
=\int \frac{\d\S_{s}(3,4)}{\d V(2)}\L(4,3;1)\; d(34)
\nonumber \\
+\int \L(1,3;2)\D(3,4)G_{s}(4,1)\; d(34)
\nonumber \\
+\int G_{s}(1,3)\D(3,4)\L(4,1;2)\; d(34),
\label{fxce}
\end{eqnarray}
where
$$
\D(1,2)=\S_{s}(1,2)-\d(1,2)v_{\xc}(1).
$$
When the potential $v_{\xc}$ has been obtained from Eq. (\ref{lss}),
the right-hand side of Eq. (\ref{fxce}) is a calculable expression for
any given approximate $\F$ and no self-consistency is required. 
As an additional bonus, all occurring Green functions are non-interacting
as opposed to interacting as one would have in most
iterative schemes based on MBPT. (Consider, {\em e.g.}, the response
function of the time-dependent Hartree-Fock approximation.)

\subsection{The exchange-only approximation}

\begin{figure}[htbp]
\begin{center}
\includegraphics*[scale=0.4]{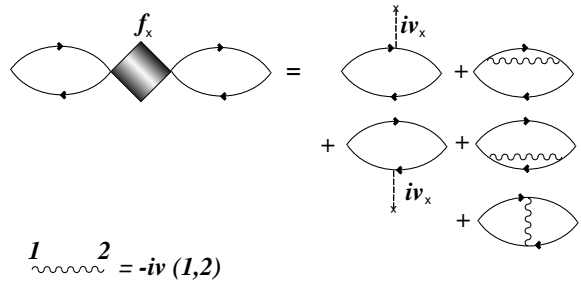}
\caption{Exchange-correlation kernel in the EXX approximation}
\label{exxk}
\end{center}
\end{figure}

Let us consider, for instance, the simplest approximation for
$\F$, namely the Hartree-Fock approximation:
$$
 \F_{\rm x} = \frac{i}{2} \Tr \{G G v\}
$$
from which we obtain
$$
\S_{\rm x}(1,2) = \frac{\d \F_{\rm x} }{\d G(2,1)} = iv(1,2)G_{s}(1,2).
$$
In this case
$$
\frac{\d\S_{\rm x}(1,2)}{\d V(3)} = iv(1,2)G_{s}(1,3)G_{s}(3,2),
$$
and Eq. (\ref{fxce}) leads to the diagrammatic expansion in Fig. 
\ref{exxk}. This approximation is also known as the
{\em exchange-only} (EXO) approximation or sometimes the {\em exact 
exchange} (EXX) approximation. It has been evaluated earlier by
several people.\cite{pgg,kvB,kg,sh,gt1,ts,has}

\subsection{The $GW$ approximation within TDDFT}

\begin{figure}[htbp]
\begin{center}
\includegraphics*[scale=0.4]{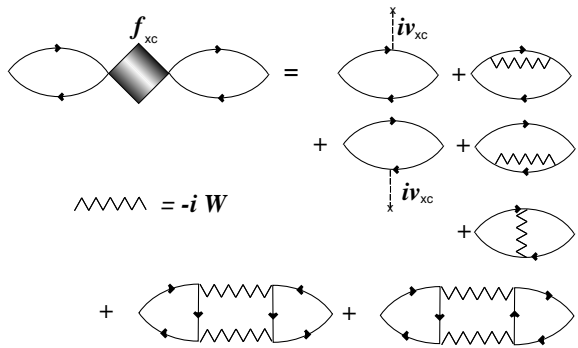}
\caption{Exchange-correlation kernel in the $GW$ approximation}
\label{gwk}
\end{center}
\end{figure}

Let us now go one level higher in the expansion of many-body 
perturbation theory and include all the screening diagrams. This is 
called the $GW$ level with $\F$ given by
$$
\F[G_{s}]=\frac{1}{2}\Tr\left\{\ln(1 + i v G_s G_s )\right\}.
$$
The expression of the self-energy in the $GW$ approximation becomes
$$
\S_{s}(1,2)=iG_{s}(1,2)W(1,2);\quad
W=[1-v\chi_{s}]^{-1}v.
$$
To calculate the variational derivative of the self-energy with 
respect to $V$ we need to evaluate the change in the screened 
potential $W$. This can  easily be constructed by observing that 
$W^{-1}=v^{-1}-\chi_{s}$ and that
$\d W/\d V=-W[\d W^{-1}/\d V]W$. The final result is displayed in 
Fig. \ref{gwk} in terms of Feynman diagrams.

All Green functions are Kohn-Sham Green functions and all 
interactions are RPA screened interactions. This response function
for which the ''time-dependent $GW$ (TD$GW$) response'' would be a
descriptive name, is presently too difficult to compute in real
systems. 
Geldart and Taylor used it to investigate the effects of the static
screening properties on the electron gas.\cite{gt2} It was used by 
Langreth and Perdew\cite{lp} in the static long wave-length limit 
in order to extract gradient approximations for DFT. 
Richardson and Ashcroft\cite{ra} have published an
approximation to the TD$GW$ response of the electron gas but only at 
imaginary frequencies.
Another application of the TD$GW$ response is due to Langreth
{\em et al.} \cite{ldrshl} and deals with Van der Waals forces.
The TD$GW$ response is generally believed to be very  accurate
but the computation of the screened interaction is known to be
a bottle neck in $GW$ calculations on real solids. Unfortunately,
the TD$GW$ response contains two such complicated factors (screened 
interactions).

\subsection{TDCDFT from the Klein Functional}

In TDCDFT the density $n$ and the physical current density $\bj$
are uniquely fixed by the external vector potential $\bA_{\rm
ext}$ and the scalar potential $w$.\cite{vr1987,cv2002} The coupling to the external
fields is given by
$$
\hat{J}=\int d^3 r \;
[\bA_{\rm ext}(\br t)\cdot \bj_{p}(\br t)+\tilde{w}(\br t)n(\br t)],
$$
where $\tilde{w}=w+\bA_{\rm ext}^{2}/2$ and ${\bf j}_p$ is the paramagnetic
current operator. According to our
prescription we render the Klein functional a functional of
$j_{\m}=(n,\bj)$ by restricting the variational freedom of the Green
functions to be all those $G_{s}$'s which are non-interacting and
given by a local scalar potential and a vector-potential, $A_{\m}=(V,\bA)$.
It is convenient to consider the four-vector $\tilde A_\mu =(\tilde V,
{\bf A})$, where $\tilde V=V+{\bf A}^2/2$, as the independent
variables since the four-vector density $j_{p,\mu}=(n,{\bf j}_{p})$ is
coupled linearly to $\tilde A_\mu$.

As in the case of only density variations, the simplicity of a
non-interacting Green function again allows the Klein functional
to be written in a much more convenient form. Using similar
manipulations as in the beginning of Sec. \ref{dftkf}, we arrive at
the expression
\be
Y_{\rm K} = T_s[n,\bj] + U_{\rm H} + \int \tilde{A}_{\m} j_{p,\m} -i \F ,
\ee
where we have used the normal convention to sum over repeated indices.
Here, the functional $T_s$ for the non-interacting kinetic energy
also depends on the physical current density $\bj$ and not only
on the density $n$. As before, the $\F$-functional plays the role of
the exchange-correlation energy. We then realize that the functional
$Y_{\rm K}$ is stationary when
\bea
\tilde{V}=\tilde{w}+V_{\rm H}+v_{\xc} \quad {\rm where}&&
\quad v_{\xc} = -i\frac{\d \F}{\d n}, \label{vcdft} \\
\bA=\bA_{\rm ext}+\bA_{\xc} \quad {\rm where}&&
\quad \bA_{\xc} = -i \frac{\d \F}{\d \bj}. \label{acdft}
\eea
Let us now focus on those system with a vanishing external vector potential.
Following the same steps as led to Eq. (\ref{lss}), {\em i.e.}, the
chain rule for differentiation, we obtain the ''linearized''
Sham-Schl\"uter equation of TDCDFT,\cite{leeuwen:proc1,leeuwen:proc2}

\begin{equation}
\int \S_{s}(2,3)\L_{\m}(3,2;1) \; d(23) = 
\int \chi_{s,\m\n}(1,2)A_{\xc,\n}(2) \; d2\, .
\label{lssc}
\end{equation}
(Notice that $A_{\xc,\m}=(v_{\xc},\bA_{\xc})$ in normal four-vector
notation.)
The generalized response function $\L_{\m}$ appearing above is
defined according to
\be
i\L_{\rm o}(2,3;1)\equiv
\frac{\d G_{s}(2,3)}{\d V(1)}=G_{s}(2,1)G_{s}(1,3),
\ee
\be
i\bL(2,3;1)\equiv
\frac{\d G_{s}(2,3)}{\d \bA(1)}=
\frac{1}{2i}G_{s}(2,1)[\overrightarrow{\grad}_{1}-
\overleftarrow{\grad}_{1}]G_{s}(1,3),
\ee
and from this response function we obtain the Kohn-Sham density-density, 
current-density, and current-current response functions
from the relation below
\be
\chi_{s,\m\n}(1,2)=\frac{\d j_{\m}(1)}{\d A_{\n}(2)}.
\ee

The many-body response function 
$$
\chi_{\m\n}(1,2)=\frac{\d j_{\m}(1)}{\d A_{\rm ext,\n}(2)},
\quad A_{\rm ext,\n}=(w,\bA_{\rm ext})
$$
can be expressed in terms of the Kohn-Sham response function 
$\chi_{s,\m\n}$ through the exchange-correlation kernel $f_{\xc,\m\n}$
\be
\chi_{\m\n}=\chi_{s,\m\n}+\chi_{s,\m\r}
(f_{\xc,\r\s}+\d_{\r \rm o}\,v\,\d_{{\rm o}\s})\chi_{\s\n}
\label{chicdft}
\ee
where $f_{\xc,\m\n}=\d A_{\xc,\m}/\d j_{\n}$. In our variational scheme
the equation for $f_{\xc}$ is obtained from one further variation of 
Eq. (\ref{lssc}) with respect to the Kohn-Sham potential $A_{\m}$.
The corresponding response function $\chi_{\m\n}$ obeys
the $f$-sum rule and Ward identities\cite{Varenna:04} 
since under a gauge transformation
the scalar potential $V$ and vector potential $\bA$ change as in
the exact CDFT, namely $V\ra V + \di f/\di t$ and
$\bA\ra \bA+\grad f$. In order to prove this property we change the
external fields according to
$w\ra w+\di f/\di t$, $\bA_{\rm ext}\ra \bA_{\rm ext}+\grad f$
and we ask the question how the scalar potential $V$ and vector
potential $\bA$ change at the stationary point.
From Eqs. (\ref{vcdft}-\ref{acdft}),
it is straightforward to realize that $V\ra V + \di f/\di t$ and
$\bA\ra \bA+\grad f$ provided the exchange-correlation potentials
change according to
$v_{\xc}\ra v_{\xc}+\bA_{\xc}\cdot \grad f$ and $\bA_{\xc}\ra \bA_{\xc}$.
Taking into account that under this gauge transformation
$G_{s}(1,2)\ra e^{-if(1)}G_{s}(1,2)e^{if(2)}$, it is a matter of
very simple algebra to show that the linearized SS equation (\ref{lssc})
is gauge invariant for any $\F$-derivable self-energy.\\

\subsection{The EXO within TDCDFT}

Let us consider, 
for instance, the exchange-only approximation for the homogeneous 
electron gas. Extracting the time-ordered component of 
Eq. (\ref{chicdft}) and taking advantage of the translational
invariance of the homogeneous electron gas, we find~\cite{leeuwen:proc1}
$$
\chi_{s,\m\r}(\bbq,\w)f_{\rm x,\r\s}(\bbq,\w)\chi_{s,\s\n}(\bbq,\w)=
V_{\m\n}(\bbq,\w)+S_{\m\n}(\bbq,\w)
$$
where all quantities are time-ordered and where $V_{\m\n}$ and
$S_{\m\n}$, at zero temperature, are given by
\begin{widetext}
\begin{eqnarray}
V_{\m\n}(\bbq,\w)=\int d^{3}p\, d^{3}k\;
p_{\m}v(|\bp-\bk|)k_{\n}
&\times&
\left\{
\frac{\theta_{\bp+\bbq/2}\bar{\theta}_{\bp-\bbq/2}}
{\w-\ve_{\bp+\bbq/2}+\ve_{\bp-\bbq/2}-i\eta}-
\frac{\bar{\theta}_{\bp+\bbq/2}\theta_{\bp-\bbq/2}}
{\w-\ve_{\bp+\bbq/2}+\ve_{\bp-\bbq/2}+i\eta}
\right\}
\nonumber \\
&\times&
\left\{
\frac{\theta_{\bk+\bbq/2}\bar{\theta}_{\bk-\bbq/2}}
{\w-\ve_{\bk+\bbq/2}+\ve_{\bk-\bbq/2}-i\eta}-
\frac{\bar{\theta}_{\bk+\bbq/2}\theta_{\bk-\bbq/2}}
{\w-\ve_{\bk+\bbq/2}+\ve_{\bk-\bbq/2}+i\eta}
\right\},
\end{eqnarray}
\begin{eqnarray}
S_{\m\n}(\bbq,\w)=\int d^{3}p\; p_{\m}p_{\n}&\times& 
\left\{
\frac{\bar{\theta}_{\bp+\bbq/2}\theta_{\bp-\bbq/2}}
{(\w-\ve_{\bp+\bbq/2}+\ve_{\bp-\bbq/2}+i\eta)^{2}}-
\frac{\theta_{\bp+\bbq/2}\bar{\theta}_{\bp-\bbq/2}}
{(\w-\ve_{\bp+\bbq/2}+\ve_{\bp-\bbq/2}-i\eta)^{2}}
\right\}\nonumber \\ 
&\times&
\left\{\S_{\rm x}(\bp+\bbq/2)-\S_{\rm x}(\bp-\bbq/2)\right\}.
\end{eqnarray}
\end{widetext}
Here, we have denoted by $p_{\m}$, $k_{\m}$ the four-dimensional vectors
of components $(1,\bp)$, $(1,\bk)$, while the Heaviside step functions 
$$
\theta_{\bbq}=\theta(\ve_{\rm F}-\ve_{\bbq})\,\, \quad \mbox{and} \,\, \quad
\bar{\theta}_{\bbq}=1-\theta_{\bbq}
$$
contain the Fermi energy $\ve_{\rm F}$.

In the large $\w$ limit the sum $V_{\rm oo}+S_{\rm oo}$ goes like 
$1/\w^{4}$ and therefore $\chi_{\rm oo}=\chi_{s,\rm oo}+O(1/\w^{4})$. 
Since the residue of the second-order pole in $\chi_{s,\rm oo}$ only 
depends on the density, the approximated response function 
$\chi_{\rm oo}$ obeys the $f$-sum rule, as it should.

\subsection{Conservation laws}

As mentioned several times, the variational and $\F$-derivable approach
to TDDFT leads to density-functional approximations which preserve many
physical properties when the system is subject to external perturbations.
Of course, TDDFT being a one-electron like theory with a multiplicative
potential trivially obeys the
continuity equation and thus particle conservation for any 
approximation to exchange and correlation. The conservation of other 
quantities will however depend on the choice of such approximations.

In this subsection we will, as an example, show how momentum conservation
follows from the general formalism. 
In the one-electron like theory of TDDFT, the change of total momentum per 
unit time is simply given by $\int n \grad (w+V_{\rm H}+v_{\xc})$.
The approximation to exchange and correlation is momentum conserving provided  
$v_{\xc}$ satisfies the zero force theorem.\cite{robert} 
Designing exchange-correlation potentials that fulfill such a 
constraint is non trivial,\cite{dbg} and several well-known 
approximations are actually not conserving.\cite{kli,robert2}
Below, we show that any approximate $v_{\xc}$ generated by our 
variational approach is fully conserving.

From Sec. III, we know that the change $\d \F$ in the $\F$-functional is
just
\begin{equation}
\d\F= i \int v_{\xc}(1)\d n(1) \; d1
\label{clbr}
\end{equation}
when we change the one-body potential from $V$ to $V+\d V$.
In the variational approach {\em a la} Klein,
Eq. (\ref{clbr}) plays a similar role as the Baym construction
$\d\F=\Tr [\S\d G]$.
In order to prove the conservation of the total momentum we have
to shown that $v_{\xc}$ does not exert any force on the Kohn-Sham system.
Let us shift all coordinates by the same time dependent infinitesimal
vector $\bdel(t)$. The functional $\F$ does not change since the
interaction potential is invariant under translations. This implies that
\be
0 = \d \F = i \int v_\xc (1) \bdel ( t_1 ) \cdot \grad_1 n(1) \; d1.
\ee
One partial integration and the fact that the vector $\bdel(t)$ is
arbitrary and independent of position gives
\be
\int n(\br t) \grad v_\xc (\br t) \; d^3 r = 0.
\ee
This means that there is no contribution from exchange and correlation
to the total force applied to the system which is given by the
classical expression $\bF=-\int n\grad w$, as it should.

The proof of momentum conservation in the presence of vector potentials
and currents follows in a similar way from the corresponding result
\begin{equation}
\d\F= i \int A_{\xc,\m}(1)\d j_{\m}(1) \; d1 ,
\label{clbrc}
\end{equation}
which we obtained from the Klein functional.\\
\\

\section{LW Functional}

Let us now discuss the variational functional of Luttinger and Ward.
From Eq. (\ref{lwf}) we find
$$
 i \d Y_{\rm LW}=\Tr\left\{\left(
\frac{1}{G_{\rm H}^{-1}-\S_{s}}-G_{s}\right)(\d\S_{s} + \d V_{\rm H})\right\}.
$$
We introduce the auxiliary Green function $\tilde{G}$ according to
$$
\tilde{G}=\frac{1}{G_{\rm H}^{-1}-\S_{s}}=G_{\rm H}+G_{\rm H}\S_{s}\tilde{G},
$$
{\em i.e.}, $\tilde{G}$ represents the first iteration toward the full
self-consistent many-body Green function starting from the Kohn-Sham
Green function $G_{s}$. Writing the total potential $V$ as
$$
V=w+V_{\rm H}+v_{\xc}
$$
and eliminating $G_{\rm H}$ between $G_{s}$ and $\tilde{G}$, one obtains
$\tilde{G}=G_{s}+\tilde{G}[\S_{s}-v_{\xc}]G_{s}$, and thus
\begin{equation}
 i \frac{\d Y_{\rm LW}}{\d V(1)}=
\Tr\left\{\tilde{G}\left[\S_{s}-v_{\xc}\right]G_{s}
\left[\frac{\d\S_{s}}{\d V(1)} + \frac{\d V_{\rm H}}{\d V(1)} \right] \right\}.
\label{lwse}
\end{equation}

\begin{figure}[htbp]
\begin{center}
\includegraphics*[scale=0.68]{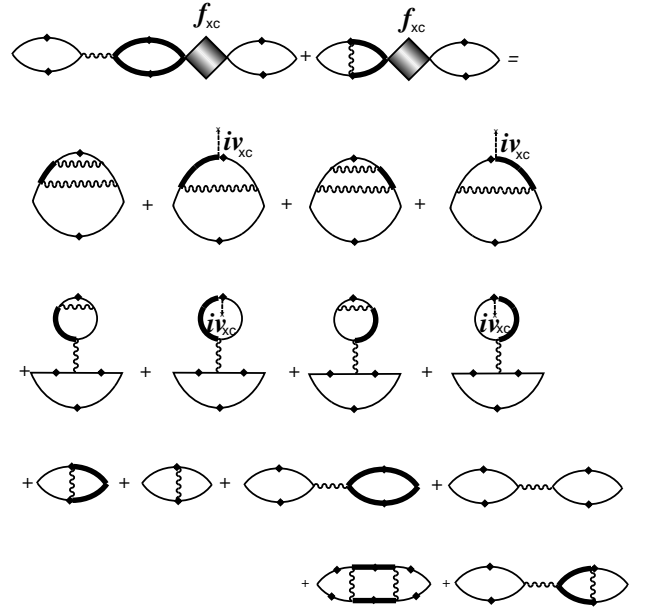}
\caption{Exchange-correlation kernel calculated from 
the LW variational approach to TDDFT in the Hartree-Fock approximation. 
The thin lines represent the noninteracting Green function $G_{s}$ 
while the tick lines represent the Green function $\tilde{G}$.}
\label{lwhffxc}
\end{center}
\end{figure}

In the Hartree-Fock approximation, $\S_{s}\simeq \S_{\rm x}$ and
Eq. (\ref{lwse}) yields
\begin{eqnarray}
\int \tilde{G}(2,3)\S_{\rm x}(3,4)G_{s}(4,5)
v(2,5)\L(5,2;1) \; d(2345)
\nonumber \\
=\int \tilde{G}(2,3)v_{\rm xc}(3)G_{s}(3,2)
v(2,4)\chi_{s}(4,1) \; d(234).
\nonumber
\end{eqnarray}
This equation determines the exchange-correlation potential which, in
turn, fixes the total potential $V$ and then $G$ and $\tilde{G}$. We
also observe that this $v_{\xc}$ is approximately linear in the
strength of the Coulomb interaction as is the ordinary exchange
potential $v_{\rm x}$ described in the context of the Klein
functional. As in the case of the EXO approximation discussed above,
the exchange correlation part of the response function $f_{\xc}$  is
obtained from one further variation with respect to the total
potential $V$. The expression now becomes slightly more tedious but
is still of the form $\hat{O}f_{\xc}=g_{\xc}$, where $\hat{O}$ is a
known operator and $g_{\xc}$ is a calculable expression which does
not involve $f_{\xc}$, see Fig. \ref{lwhffxc}.

The response function of the LW functional at the Hartree-Fock level
is expected to be superior
to that of the EXO due to the better stability of the LW functional as
compared to the Klein functional. 

We also wish to emphasize that it made no difference to the
resulting response function whether we used the LW functional or the
Klein functional to derive it - provided we allowed free variations
of the Green function $G$. By restricting the possible choices of
Green functions to those produced by local potentials (TDDFT) the LW
and Klein functionals give rise to different response functions at
the same level of many-body perturbation theory.

\section{$\Q$ Functionals}

The main advantage of the $\Q$ functionals is that they give the
possibility of using physical models for the screening, the
calculation of which is actually a bottleneck in practical
applications. A word of caution is, however, appropriate in this
context. With model screened interactions ($W$'s) there is usually
no self-consistency with respect to $W$, a fact that might compromise
the conserving property of the theory. The $\Q$ functionals have two
independent arguments ($G$ and $W$) resulting in terms linear in the
deviation of the actual Green function from the self-consistent one
when $W$ is away from the value which renders the functional stationary.
When the $\Q$ functionals are used to construct
response functions of TDDFT the theory is, however, variational with
respect to the one-body potential generating the non-interacting Green
function - even when a model $W$ is used.
This fact actually restores several conserving properties although
this has to be verified from case to case. For instance, choosing model
$W$:s which, like the bare Coulomb interaction, are instantaneous and
translationally invariant will clearly not spoil the conserving properties.

The first $\Q$ functional was constructed by ABL in 1996.\cite{abl} It has 
the appearance
\begin{eqnarray}
 i Y_{\rm ABL}[G,W]&=&\Q[G,W]-\Tr\left\{\S G-\ln\left(\S-G_{\rm H}^{-1}\right)
\right\}
\nonumber \\
&+&\frac{1}{2}\Tr\left\{
WP+\ln\left(1-vP\right)\right\}-i U_{\rm H}[G].
\nonumber \\
\label{ablf}
\end{eqnarray}
The ABL functional is stationary with respect to variations of $G$ 
and $W$ whenever $G$ obeys Dyson's equation and $W$ obeys the 
``contracted Bethe-Salpeter equation'', 
$W=v+vPW$. As for the $\F$ functionals, the self-energy is obtained 
by taking the functional derivative of $\Q$ with respect to $G$ and 
the polarization $P$ turns out to be the negative of twice the functional 
derivative of $\Q$ with respect to $W$
$$
\S(1,2)=\frac{\d\Q}{\d G(2,1)},\quad
P(1,2)=-2\frac{\d\Q}{\d W(2,1)}.
$$
Just as was the case for the pure $\F$ functionals, we can add to 
any $\Q$ functional an arbitrary functional $K[Q]$ of a quantity $Q$ 
defined by $Q=W\left(v^{-1}-P[G,W]\right)-1$, with the properties
$$
K[0]=\frac{\d K}{\d Q}[0]=0 .
$$
We then obtain a new $\Q$ functional with the same stationary point and
the same value at the stationary point. An example of a simple 
functional obtained in this way is
\begin{eqnarray}
iY_{\rm LWS}[G,W]=\Q[G,W]-
\Tr\left\{\S G-\ln\left(\S-G_{\rm H}^{-1}\right)\right\}
\nonumber \\
+\frac{1}{2}\Tr\left\{
Wv^{-1}-1+\ln\left(Wv^{-1}\right)\right\}-iU_{\rm H}[G].
\nonumber
\end{eqnarray}
Here, LWS stands for the simple version of the $\Q$ functional based on
the construction of Luttinger and Ward.
As for the Klein version of $\F$-derivable functionals, we expect 
that this functional is less stable than the original ABL functional 
of Eq. (\ref{ablf}). Also, it contains the same LW expression
which led to the complicated result depicted in Fig. \ref{lwhffxc} 
and, for the moment, we deem this functional as less suitable for
the construction of response functions.

Another possibility is to, instead, add to $Y_{\rm ABL}$ 
the functional $F[D]$ of Eq. (\ref{fdfunc}), thus obtaining the Klein 
version of $\Q$-derivable functionals
\begin{eqnarray}
iY_{\rm ABLK}[G,W]&=&\Q[G,W]
\nonumber \\ 
&-&\Tr\left\{GG_{\rm H}^{-1}-1+\ln\left(-G^{-1}\right)\right\}
\nonumber \\
&+&\frac{1}{2}\Tr\left\{
WP+\ln\left(1-vP\right)\right\}-i U_{\rm H}[G].
\nonumber
\end{eqnarray}
Again, due to the simplicity of the ''Klein'' expression, we can here
use the same manipulations as we applied to the original Klein 
functional in order to arrive at Eq. (\ref{kleinxc}). Thus, inserting
the non-interacting Green function $G_s$ into the functional
$Y_{\rm ABLK}$, we then obtain
\be
Y_{\rm ABLK}[V] = T_s[n] + \int w n + U_{\rm H} + E_{\xc}[n] ,
\ee
where
\begin{eqnarray}
E_{\xc}[n] &=& -i \Q[G_s,W] \nonumber \\
&-&\frac{i}{2}\Tr\left\{W P[G_s,W] + \ln\left(1-vP[G_s,W]\right)\right\}.
\nonumber
\end{eqnarray}
Consequently, also this functional can be given the standard DFT form and
we realize that it is stationary at the non-interacting Green
function $G_s$ produced by the local one-electron potential
\be
V = w + V_{\rm H} + \frac{\d E_{\xc}}{\d n} .
\label{vtot}
\ee
In fact, all functionals, be they of the $\F$ or the $\Q$ variety, 
having the ``Klein'' form for their dependence on the external 
potential $w$ have the nice property that the optimizing potential 
consists of the external potential $w$, the Hartree potential 
$V_{\rm H}$, and the functional derivative of the exchange-correlation 
energy with respect to the density $n$. In Eq. (\ref{vtot}), the last 
derivative is calculated from the chain rule for differentiation 
giving the OPM-like equation 
\begin{eqnarray}
&&\int \chi_{s}(1,2)v_{\xc}(2) \; d2 = \frac{\d E_{\xc}}{\d V(1)} =
\int \L(3,2;1) 
\nonumber \\ 
&&\times\left[\S(2,3)+\frac{1}{2}\int\D W(4,5)\frac{\d P(5,4)}{\d G_{s}(3,2)}
\;d(45)\right]
\; d(23) .
\nonumber
\end{eqnarray}
The quantity $\D W$ is $W-\tilde{W}=W-v/(1-vP)$ and we remind the 
reader that we are here allowed to use any model for $W$. In 
particular, we could choose $W$ to be $\tilde{W}$, in which case our 
equation for the exchange-correlation potential $v_{\xc}$ reduces to 
the same expression as obtained from the ``Klein version'' of the 
$\F$ formalism described in Sec. \ref{dftkf}. Furthermore, it is 
easily seen directly from its definition that the functional 
$Y_{\rm ABLK}$ becomes independent of the choice of model $W$ at the 
level of the RPA. Thus, at that level, this functional does not add 
anything to the previously discussed $\F$-derivable scheme at the 
same level (RPA). Being, for the moment, content with that level we 
will here not pursue the $Y_{\rm ABLK}$ any further.

\begin{figure}[htbp]
\begin{center}
\includegraphics*[scale=0.4]{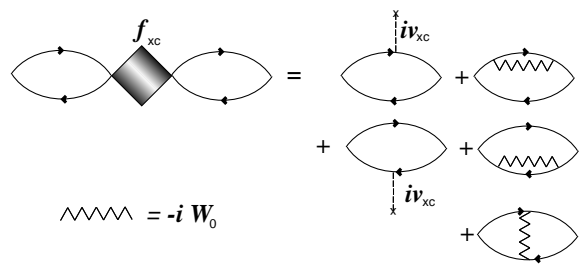}
\caption{Exchange-correlation kernel at the $GW$ level in the $\Q$ 
scheme.}
\label{gwkabl}
\end{center}
\end{figure}

Finally, by adding an appropriate choice for the functional $K[Q]$, as 
discussed above, to the functional $Y_{\rm ABLK}$, we obtain the 
simplest functional $Y_{\rm KK}$ of those discussed in the present work. 
We have
\begin{eqnarray}
iY_{\rm KK}&=&\Q-
\Tr\left\{GG_{\rm H}^{-1}-1+\ln\left(-G^{-1}\right)\right\}
\nonumber \\ 
&+&\frac{1}{2}\Tr\left\{
Wv^{-1}-1+\ln\left(Wv^{-1}\right)\right\}-iU_{\rm H}[G].
\nonumber
\end{eqnarray}
As before, we restrict the variational freedom of the Green functions
to non-interacting ones ($G_{s}$) and differentiate $Y_{\rm KK}$ 
with respect to the total, as it turns out,  Kohn-Sham
potential $V$ producing the non-interacting $G_{s}$.
Notice that $W$ is an independent variable and does not depend on 
$G_{s}$ or $V$. Only $G_{s}$ depends on $V$. We obtain
\begin{equation}
\int
\S_{s}(2,3)\L(3,2;1) \; d(23) =
\int \chi_{s}(1,2)v_{\xc}(2) \; d2 ,
\label{lssq}
\end{equation}
where $v_{\xc}=V-w-V_{\rm H}$, as before. We are now allowed
to choose any appropriate but approximate $W=W_{\rm o}$.
Let us study the response function resulting from the functional
$Y_{\rm KK}$ at the $GW$ level, meaning that
$$
\Q=\frac{i}{2}\Tr\left\{G_{s}G_{s}W_{\rm o}\right\}.
$$
Consequently, in this approximation, we obtain the $GW_{\rm o}$
self-energy
$$
\S_{s}=\frac{\d \Q}{\d G_{s}}=iG_{s}W_{\rm o}.
$$
One further variation of
Eq. (\ref{lssq}) with respect to $V$ gives an equation for $f_{\xc}$,
whose diagrammatic representation is given in Fig. \ref{gwkabl}.
Here, the Green functions are Kohn-Sham Green functions and the
corresponding exchange-correlation potential is that which, to first
order, reproduces the density of a $GW$ calculation with some model
$W_{\rm o}$. The screened interactions could be model interactions,
{\em e.g.}, Yukawa potentials, screened potentials within the RPA,
or plasmon-pole approximations.
The vertex diagram in Fig. \ref{gwkabl}
should be calculated with Kohn-Sham Green functions. The sum of the
first four self-energy diagrams is essentially, to first order,
equivalent to a single polarization diagram calculated using
$GW_{\rm o}$ Green functions.

If $W_{\rm o}$ is chosen to be a Yukawa potential, {\em i.e.}, a
statically screened Coulomb interaction, this conserving $GW_{\rm o}$
response function provides a justification of the work of Marini {\em
et. al.}\cite{mr} provided the Green function with $GW$ (RPA) shifted
poles is close to the Green function of a Hartree-Fock calculation
with a statically screened Coulomb potential.

The use of a Yukawa potential for a screened interaction also sheds
some light on the work by Kim and G\"orling.\cite{kg} They did exactly
the TDDFT response function of Fig. \ref{gwkabl} in bulk silicon using
the bare Coulomb interaction for $W$. They found that their calculated
optical absorption spectrum was far from an experimental result unless
they, somewhat artificially, cut down the range of the
particle-hole interaction.

It would actually be quite interesting to investigate the properties of
the response function of Fig. \ref{gwkabl} using a variety of screened
interactions of the kind that preserves its conserving
properties. An ongoing collaborative project\cite{nq,ar_mg} has precisely
this objective in mind.

\section{Conclusions and Outlooks}

In the present work we have proposed a new way of obtaining approximations
to current and density response functions of realistic systems. Our theory
is based on the variational approach to many-body theory as previously
formulated by us and others. It gives the possibility to find successively
better approximations to the effects of exchange and correlation in the
framework of time-dependent density-functional or current density
functional theory.  The fact that the theory is formulated in the language
of TDDFT makes it much easier, from a computational perspective, to apply
to realistic systems as compared to standard MBPT.

Improved approximations can be constructed in a systematic way
in the same sense as within MBPT. But physical intuition as to what
physical processes are important for any particular problem must be
applied. The underlying variational approach to MBPT is not unique and
many variational functionals can be constructed leading to the same
quality of approximation. Our method for improved approximations
within TDDFT has the same feature. Different functionals have different
variational accuracy meaning different sizes of the second order errors.
In the present paper we have discussed mainly two functionals - that due
to Luttinger and Ward (LW) and that due to Klein (K). The former has
proved to be more stable as compared to the latter as far as
concerns the calculation of total energies of a variety of systems
ranging from those with very localized electrons to those with itinerant
electrons. This would suggest that the LW functional ought to be used
also for the construction of response functions within TDDFT. In the
present work, we have given the formulas for the exchange-correlation
kernel of TDDFT resulting from both functionals. Sadly enough, we judge
that of the supposedly better LW functional to be beyond our present
computational facilities - even at a rather low level of approximation
within MBPT. In order to demonstrate this point, we have given the diagrams
representing the density response function resulting from the LW
formulation within the exchange-only approximation.
We would still like to draw the readers attention to the fact that the
ambiguity in the choice of functionals can most likely be used to our
advantage. But much more research is needed in order to see how
this should be done.

A very important feature of our variational approach to TDDFT is the
fact it relies on the $\F$ or $\Q$ derivability of the underlying
approximation within MBPT. Combined with the variational property of
the chosen functional, this leads to the preservation of many physically
important conservation laws and sum rules. And this is true regardless
of the actual chosen level of approximation within MBPT. This highly
desirable feature is not guaranteed in other available approaches
based on straight-forward diagrammatic expansions, iterative techniques,
or decoupling schemes. 
For instance, in Ref. \onlinecite{punk} one develops a diagrammatic 
representation for the particular many-body perturbation scheme which 
starts from a zero-th order Hamiltonian which already gives the 
correct density. \cite{gl} Unfortunately, this technique suffers from 
the same basic drawbacks as ordinary MBPT - it is, in principle,
divergent, summations must be carried to infinite order, and
there is no guarantee for obtaining approximations which have certain 
desirable physical properties automatically included. The same holds 
true for expansions which are based on iterating the so called 
Hedin equations\cite{hedin} using the screened interaction as the 
``small'' parameter.\cite{lucia}
As an example, we have, in the present work, demonstrated how
the variational approach leads to momentum conservation in the case of
the Klein functional.

It is worth while observing that the so called "linearized"
Sham-Schl\"uter equation actually turns out to be a result of our
variational approach starting from the Klein functional. But this is
only true if the self-energy involved is a $\F$- or $\Q$-derivable one.
In that case, the resulting approximation for the response function is,
of course, conserving.

We also remark that the so called optimized potential method (OPM) and
many generalizations thereof readily follows from the theory presented
here. As an example, we have given the explicit formulas for the current
density response of a homogeneous system within the exchange-only
approximation.

Even though we now have a systematic way of obtaining better response
functions within TDDFT the expressions quickly become too complicated
to be implemented in low-symmetry systems, especially when we want to
include all physically relevant processes. In this context, we advocate
the use of the $\Q$-derivable theories which allows for the use of model
screened interactions without loosing the important conserving properties.
In this way, important physical effects like, {\em e.g.}, a strong
particle-hole interaction can be incorporated without an excessive
increase in the computational effort. One should, however, keep in
mind that models for the screened interaction must possess certain
symmetries related to the actual system in order for the conserving
properties to be preserved.

We have discussed the implementation of the theories presented here with
other research groups. One particularly promising approach is that which
is based on the Klein functional and the $\Q$-formulation using a model
screened interaction like, {\em e.g.}, a statically screened Coulomb
interaction as often used in the Bethe-Salpeter approach, or a simple
plasmon-pole approximation. Together with our collaborators,\cite{nq,ar_mg}
we hope to
be able to present some numerical results within the near future.

\section*{ACKNOWLEDGEMENTS}

We would like to thank Carl-Olof Almbladh for fruitful discussions
during the course of this work.

This work was supported by the European Cummunity 6:th framework Network of
Excellence NANOQUANTA (NMP4-CT-2004-500198).

\end{document}